\documentclass[a4paper,11pt]{article}

\usepackage{jheppub}
\usepackage{graphicx}
\usepackage{amsmath}
\usepackage{slashed}
\usepackage{braket}
\usepackage[export]{adjustbox}
\usepackage{enumitem}
\usepackage{placeins}
\usepackage[normalem]{ulem}

\newcommand{\refcite}[1]{ref.~\cite{#1}}
\newcommand{\refscite}[1]{refs.~\cite{#1}}
\newcommand{\Eq}[1]{Eq.~\eqref{eq:#1}}
\newcommand{\eq}[1]{eq.~\eqref{eq:#1}}
\newcommand{\eqs}[2]{eqs.~\eqref{eq:#1} and \eqref{eq:#2}}
\renewcommand{\sec}[1]{section~\ref{sec:#1}}

\newcommand{\app}[1]{appendix~\ref{app:#1}}
\newcommand{\fig}[1]{figure~\ref{fig:#1}}

\newcommand{\df}{\mathrm{d}}
\newcommand{\img}{\mathrm{i}}
\newcommand{\eps}{\epsilon}

\newcommand{\xb}{\bar x}
\newcommand{\bn}{{\bar n}}

\newcommand{\GeV}{\,\mathrm{GeV}}
\newcommand{\cI}{\mathcal{I}}

\newcommand{\cL}{\mathcal{L}}
\newcommand{\cM}{\mathcal{M}}
\newcommand{\cN}{\mathcal{N}}
\newcommand{\cO}{\mathcal{O}}

\newcommand{\qt}{{\vec q}_T}
\newcommand{\bt}{{\vec b}_T}

\newcommand{\wa}{{w_1}}
\newcommand{\wb}{{w_2}}

\newcommand{\nlim}{\lim\limits_{\mathrm{strict}~n-\mathrm{coll.}}}
\newcommand{\as}{\alpha_s}
\newcommand{\GammaC}{\Gamma_{\rm cusp}}
\newcommand{\nn}{\nonumber}
\newcommand{\MSbar}{\overline{\mathrm{MS}}}
\newcommand{\lqcd}{\Lambda_\mathrm{QCD}}

\def\beq{\begin{equation}}
\def\eeq{\end{equation}}
\def\bea{\begin{eqnarray}}
\def\eea{\end{eqnarray}}

\title{\boldmath Transverse momentum dependent PDFs at N$^3$LO}

\author[a]{Markus A.~Ebert,}
\emailAdd{ebert@mit.edu}

\author[b]{Bernhard Mistlberger,}
\emailAdd{bernhard.mistlberger@gmail.com}

\author[a]{and Gherardo Vita}
\emailAdd{vita@mit.edu}

\affiliation[a]{Center for Theoretical Physics, Massachusetts Institute of Technology, Cambridge, Massachusetts 02139, USA}
\affiliation[b]{SLAC National Accelerator Laboratory, Stanford University, Stanford, CA 94039, USA}

\abstract{
We compute the quark and gluon transverse momentum dependent parton distribution functions at next-to-next-to-next-to-leading order (N$^3$LO) in perturbative QCD. 
Our calculation is based on an expansion of the differential Drell-Yan and gluon fusion Higgs production cross sections about their collinear limit.
This method allows us to employ cutting edge multiloop techniques for the computation of cross sections to extract these universal building blocks of the collinear limit of QCD.
The corresponding perturbative matching kernels for all channels are expressed in terms of simple harmonic polylogarithms up to weight five.
As a byproduct, we confirm a previous computation of the soft function for transverse momentum factorization at  N$^3$LO.
Our results are the last missing ingredient to extend the $q_T$ subtraction methods to N$^3$LO and to obtain resummed $q_T$ spectra at N$^3$LL$^\prime$ accuracy both for gluon as well as for quark initiated processes.
}

\preprint{MIT-CTP 5209, SLAC-PUB-17535}

\begin{document}

\maketitle

\section{Introduction}
\label{sec:intro}

Transverse momentum dependent parton distribution functions (TMDPDFs) extend the concept of collinear PDFs,
which describe the longitudinal momentum distribution of quarks and gluons inside protons,
to also reflect their intrinsic transverse motion.
They are important ingredients for describing high-energy scatterings at small transverse momentum,
in particular the Drell-Yan process, an important benchmark observable of the Standard Model
both at the Tevatron~\cite{Affolder:1999jh, Abbott:1999yd, Abazov:2007ac, Abazov:2010kn}
and the LHC~\cite{Aad:2011gj,Chatrchyan:2011wt,Aad:2014xaa,Khachatryan:2015oaa,Aad:2015auj,Khachatryan:2016nbe,Sirunyan:2019bzr,Aad:2019wmn}.
Similarly, they are required for predictions of the Higgs transverse momentum spectrum,
a key observable that is of great interest for the LHC physics program
\cite{Aad:2014lwa,Aad:2014tca,Aad:2016lvc,Aaboud:2017oem,Aaboud:2018xdt,ATLAS:2020wny,Khachatryan:2015rxa,Khachatryan:2015yvw,Khachatryan:2016vnn,Sirunyan:2018kta,Sirunyan:2018sgc}.
TMDPDFs also arise in measurements of semi-inclusive deep-inelastic scattering (SIDIS) \cite{Ashman:1991cj,Derrick:1995xg,Adloff:1996dy,Aaron:2008ad,Airapetian:2012ki,Adolph:2013stb,Aghasyan:2017ctw},
where they are of particular interested because they provide a window into the proton structure~\cite{Boer:2011fh,Accardi:2012qut}.

TMDPDFs measure both the longitudinal momentum fraction $z$ and the transverse momentum $\qt$ carried by the struck parton.
They are intrinsically nonperturbative objects that need to be extracted from measurements,
but for perturbative $|\qt|$ they can be perturbatively related to collinear PDFs.
Schematically, this matching takes the form
\begin{align} \label{eq:beam_OPE_schematic}
 B_i(z, \qt) = \sum_j \int_z^1\!\frac{\df z'}{z'} \, \cI_{ij}(z',\qt) f_j\Bigl(\frac{z}{z'}\Bigr) \times \bigl[1 + \cO(q_T/\lqcd)\bigr]
\,,\end{align}
where $B_i$ is the so-called TMD beam function for a parton of flavor $i$, the sum runs over all parton flavors $j$, $\cI_{ij}$ is the perturbative matching kernel, and $f_j$ is the collinear PDF.
The precise knowledge of $\cI_{ij}$ is important for measurements dominated by transverse momenta that are small compared to the hard scale $Q$ of the process but still perturbative, i.e.\ $\lqcd \ll |\qt| \ll Q$, such as Higgs and Drell-Yan production at the LHC. Precise perturbative predictions for the beam function are also essential to extract the intrinsically nonperturbative corrections, which in global fits is typically achieved through a nonperturbative model on top of \eq{beam_OPE_schematic}, see e.g.~\refscite{Landry:1999an,Landry:2002ix,Konychev:2005iy,DAlesio:2014mrz,Bacchetta:2017gcc,Scimemi:2017etj,Scimemi:2019cmh}.

Since TMDPDFs describe processes at small transverse momentum $\qt$, they are intrinsically sensitive to the infrared structure of QCD.
Hence, their perturbative structure is intimately related to the singular structure of QCD amplitudes.
This property is employed in the $q_T$ subtraction scheme proposed in \refcite{Catani:2007vq}
to achieve the cancellation of infrared divergences in next-to-next-leading order (NNLO) calculations of color-singlet cross sections.
Recently, extensions of this method to next-to-next-to-next-to-leading order (N$^3$LO) were discussed in \refscite{Cieri:2018oms,Billis:2019vxg}, which however did not include the required three-loop ingredient.

TMDPDFs are composed of the TMD beam function $B_i(z,\qt)$ and the TMD soft function $S(\qt)$.
The soft function has been known at three loops since quite some time~\cite{Echevarria:2015byo,Li:2016ctv,Luebbert:2016itl},
and the quark beam function has been calculated at this accuracy recently~\cite{Catani:2012qa,Gehrmann:2012ze,Gehrmann:2014yya,Echevarria:2016scs,Luo:2019hmp,Luo:2019szz},
while the gluon beam function so far is only known at two loops~\cite{Catani:2011kr,Gehrmann:2014yya,Echevarria:2016scs,Luo:2019bmw}.
In this paper, we fill this gap by calculating the full matching coefficient for the gluon beam function at N$^3$LO.
We also calculate the full quark beam function at N$^3$LO, where we find disagreement with the recent calculation of the corresponding result in \refcite{Luo:2019szz} in the $d_{abc} d^{abc}$ color structure.
These results make it possible to fully apply the $\qt$ subtraction at N$^3$LO accuracy,
paving the way for fully-differential cross sections of color-singlet processes at this order.
Our results are also the last missing ingredient for TMD resummation at N$^3$LL$^\prime$ accuracy.
They also arise in $\qt$-dependent event shapes at hadron colliders such as the Transverse Energy-Energy Correlator (TEEC)~\cite{Gao:2019ojf}, and for the azimuthal angle in vector boson $+j$ production in the back-to-back limit~\cite{Chien:2019gyf, Chien:2020hzh}.

We perform the calculation of the TMDPDF at N$^3$LO by using the framework of the collinear expansion of cross sections presented in~\cite{Ebert:2020lxs}.
This framework allows us to efficiently compute universal building blocks of perturbative QFT in kinematic limits leveraging on modern technology developed for the computation of multiloop scattering cross sections.
In particular, we expand the diagrams for the Drell-Yan and gluon fusion Higgs boson production cross section at N$^3$LO in the collinear limit.
We make use of the framework of reverse unitarity~\cite{Anastasiou2003,Anastasiou:2002qz,Anastasiou:2003yy,Anastasiou2005,Anastasiou2004a} to enforce measurement and on-shellness constraints on the final states as well as integration-by-part (IBP) identities~\cite{Chetyrkin:1981qh,Tkachov:1981wb} and the method of differential equations~\cite{Kotikov:1990kg,Kotikov:1991hm,Kotikov:1991pm,Henn:2013pwa,Gehrmann:1999as} to obtain the cross sections differential in the rapidity and transverse momentum of the colorless final states in the collinear limit.
Following \refcite{Ebert:2020lxs}, we exploit this limit of the differential cross sections to extract the bare matching kernels of the quark and gluon $\qt$ beam functions.

This paper is structured as follows. In \sec{factorization}, we briefly review TMD factorization. 
In \sec{coll_exp}, we discuss how the beam function can be calculated from the collinear limit of a color-singlet cross section using the method collinear expansions.
In \sec{results}, we present our results, before concluding in \sec{conclusion}.
Our results are also available in electronic form in the ancillary files of this submission.

\section{Review of \texorpdfstring{\boldmath $q_T$}{qT} factorization}
\label{sec:factorization}

We study the production of a color-singlet state $h$
and an additional hadronic state $X$ in a proton-proton scattering process,
\begin{align} \label{eq:process_hadr}
 P(P_1) + P(P_2) \quad\to\quad h(-p_h) + X(-k)
\,,\end{align}
where we align the incoming protons along the directions
\begin{align} \label{eq:nnb}
 n^\mu = (1,0,0,1) \,,\qquad \bn^\mu = (1,0,0,-1)
\end{align}
and denote their momenta as $P_1$ and $P_2$, with the center of mass energy being $S=(P_1+P_2)^2$.
We are interested in measuring the cross section differential in $p_h^\mu$,
expressed through the invariant mass $Q^2 = p_h^2$, rapidity $Y$, and transverse momentum $\qt$.

The factorization of the cross section in the limit $q_T \ll Q$ was first established by Collins, Soper, and
Sterman (CSS)~\cite{Collins:1981uk,Collins:1981va,Collins:1984kg},
and was further elaborated on in \refscite{Catani:2000vq,deFlorian:2001zd,Catani:2010pd,Collins:1350496}.
The factorization was also shown using Soft-Collinear Effective Theory (SCET)
\cite{Bauer:2000ew, Bauer:2000yr, Bauer:2001ct, Bauer:2001yt}
by several groups \cite{Becher:2010tm,Becher:2011xn,Becher:2012yn,GarciaEchevarria:2011rb,Echevarria:2012js, Echevarria:2014rua,Chiu:2012ir,Li:2016axz}.
The factorized cross section is typically expressed in Fourier space in two equivalent forms,
\begin{align} \label{eq:TMD_factorization}
 \frac{\df\sigma}{\df Q^2 \df Y \df^2 \qt} &
 = \sigma_0 \sum_{a,b} H_{ab}(Q^2,\mu) \int\!\frac{\df^2\bt}{(2\pi)^2} e^{\img\,\qt \cdot \bt}
   \,\tilde B_a\Bigl(x_1^B, b_T, \mu, \frac{\nu}{\omega_a}\Bigr)
   \,\tilde B_b\Bigl(x_2^B, b_T, \mu, \frac{\nu}{\omega_b}\Bigr)
   \, \tilde S(b_T, \mu, \nu)
\nn\\&
 = \sigma_0 \sum_{a,b} H_{ab}(Q^2,\mu) \int\!\frac{\df^2\bt}{(2\pi)^2} e^{\img\,\qt \cdot \bt}
   \,\tilde f_a(x_1^B, b_T, \mu, \zeta_a) \, \tilde f_b(x_2^B, b_T, \mu, \zeta_b)
\,.\end{align}
For processes inclusive in $h$, \eq{TMD_factorization} holds up to corrections in $\cO(q_T^2/Q^2)$. Power corrections of $\cO(q_T^2/Q^2)$ have been firstly calculated at fixed order in perturbation theory in \refcite{Ebert:2018gsn}, while the study of their all order structure has been initiated using SCET operator formalism~\cite{Kolodrubetz:2016uim,Feige:2017zci,Moult:2017rpl,Chang:2017atu,Chang:NLP} and their nonperturbative structure has been explored in \refscite{Balitsky:2017gis,Balitsky:2017flc}.
\Eq{TMD_factorization} receives enhanced $\cO(q_T/Q)$ corrections when applying fiducial cuts to $h$ \cite{Ebert:2019zkb} that can be uniquely included in the factorization for Higgs and Drell-Yan production~\cite{Ebert:fiducial}, and also receives linear corrections when radiation from massive final states is considered \cite{Buonocore:2019puv}.

In \eq{TMD_factorization}, we sum over all parton flavors $a$ and $b$ mediating
the underlying partonic process $ab \to h$. The corresponding partonic
Born cross section is denoted by $\sigma_0$, and the hard function $H_{ab} = 1 + \cO(\as)$
encodes virtual corrections to the Born process.
For Drell-Yan and gluon fusion Higgs production in the $m_t\to\infty$ limit, the N$^3$LO hard function can be found in \refcite{Gehrmann:2010ue}, and for $b \bar{b} \to H$ in \refscite{Gehrmann:2014vha,Ebert:2017uel}.
The $\tilde B_i(x,b_T,\mu,\nu/\omega)$ encode the probability to find a parton of flavor $i$
with longitudinal momentum fraction $x$ and impact parameter $\bt$, which is Fourier-conjugate to $\qt$.
The soft function $\tilde S$ encodes the effect of soft radiation from either proton.
In the second line of \eq{TMD_factorization}, these functions are combined into the TMDPDF
\beq\label{eq:TMDPDFdef}
 \tilde f_i(x, b_T, \mu, \zeta) = \tilde B_i\Bigl(x, b_T, \mu, \frac{\nu}{\sqrt\zeta}\Bigr) \sqrt{S(b_T, \mu, \nu)}
\,,\eeq
which is indepdent of the rapidity scale $\nu$ discussed below.
Computationally, it is natural to separately consider
the calculation of the beam and soft functions appearing in \eq{TMD_factorization},
which can easily be combined into the TMDPDF if desired.

A characteristic feature of $q_T$ factorization is the appearance of so-called
rapidity divergences~\cite{Collins:1981uk,Collins:2008ht,Becher:2010tm,GarciaEchevarria:2011rb,Chiu:2011qc,Chiu:2012ir,Rothstein:2016bsq},
which require an explicit rapidity regulator. Similar to the emergence of the
renormalization scale $\mu$ from ultraviolet (UV) regularization, this induces a rapidity scale,
which we generically as $\nu$. The rapidity divergences track the energy of the
struck partons, encoded in the parameters
\begin{align}
 \omega_a = x_a \, \bn \cdot P_a = Q e^{+Y}
\,,\quad
 \omega_b = x_b \, n \cdot P_b = Q e^{-Y}
\,,\qquad
 \zeta_{a,b} \propto \omega_{a,b}^2\quad\mathrm{s.t.}\quad\mathrm\zeta_a \zeta_b = Q^4
\,.\end{align}
There is a variety of rapidity regulators, giving rise to several formulations
of the individual ingredients in \eq{TMD_factorization}. However, all approaches
yield the same fixed-order results for the physical cross section in \eq{TMD_factorization}.
Thus, we are free to choose the regulator most convenient for our calculation,
and we will employ the exponential regulator of \refcite{Li:2016axz}.
Explicit definitions of the beam and soft functions in terms of gauge-invariant
matrix elements formulated in SCET can be found in \refcite{Li:2016axz}
(see also \refscite{Luo:2019hmp,Luo:2019bmw}), but are not required in our approach.

In this work, we focus on TMD factorization in the perturbative regime $b_T^{-1} \sim q_T \gg \lqcd$,
in which the TMD beam function and TMDPDF can be matched onto PDFs as \cite{Collins:1984kg}
\begin{align} \label{eq:beam_OPE}
 \tilde B_i\Bigl(z, b_T, \mu, \frac{\nu}{\omega}\Bigr) &
 = \sum_j \int_z^1\!\frac{\df z'}{z'} \, \tilde \cI_{ij}\Bigl(z', b_T, \mu, \frac{\nu}{\omega}\Bigr)
   f_j\Bigl(\frac{z}{z'},\mu\Bigr) \times \bigl[1 + \cO(b_T \lqcd)\bigr]
\,,\nn\\
 \tilde f_i(z, b_T, \mu, \zeta) &
 = \sum_j \int_z^1\!\frac{\df z'}{z'} \, \tilde\cI^{\rm TMD}_{ij}(z', b_T, \mu, \zeta)
   f_j\Bigl(\frac{z}{z'},\mu\Bigr) \times \bigl[1 + \cO(b_T \lqcd)\bigr]
\,.\end{align}
The matching kernels $\tilde\cI_{ij}$ and $\tilde\cI^{\rm TMD}_{ij}$ are the objects of interest of this paper.

\section{Beam functions from the collinear limit of cross sections}
\label{sec:coll_exp}

We consider the contribution to \eq{process_hadr} from the partonic process
\begin{align} \label{eq:process_part}
 i(p_1) + j(p_2) \quad\to\quad h(-p_h) + X_n(-p_3, \dots, -p_{n+2})
\,.\end{align}
The incoming partons carry momentum $p_1$ and $p_2$ and flavor $i$ and $j$, respectively, while we denote with $X_n$ the hadronic final state with total momentum $k$, consisting of $n$ partons with momenta $\{-p_3, \dots, -p_{n+2}\}$, such that $k = \sum_{i\ge3} p_i$. Note that at tree level we have $n=0$.

The final-state momenta are parameterized in terms of
\begin{align} \label{eq:vardef}
 Q^2 = p_h^2
 \,, \quad
 Y = \frac12 \ln\frac{\bn \cdot p_h}{n \cdot p_h}
 \,, \quad
 \wa = -\frac{\bar n\cdot k}{\bar n \cdot p_1}
 \,,\quad
 \wb = -\frac{ n\cdot k}{ n \cdot p_2}
 \,,\quad
 x = \frac{k^2}{(\bar n\cdot k)(n\cdot k)}
\,,\end{align}
where $Y$ is the rapidity of the color-singlet state $h$ and $Q^2$ its invariant mass.

The partonic cross section differential in these variables is defined as
\begin{align} \label{eq:sigma_part}
  \frac{\df \eta_{ij}}{ \df Q^2  \df \wa \df \wb \df  x}  &
 = \frac{1}{\sigma_0} \frac{\cN_{ij}}{2 S} \sum_{X_n}\int  \frac{\df\Phi_{h+n} }{ \df \wa \df \wb \df  x}\, |\cM_{ij\to h+X_n}|^2
\,.\end{align}
Here $\df\Phi_{h+n}$ represents the differential phase space measure for the $h+X_n$ state,
$|\cM_{ij\to h+X_n}|^2$ is the squared matrix element for the partonic process in \eq{process_part},
summed over the colors and helicities of the particles, with $\cN_{ij}$ containing the helicity and color average of the incoming particles, and we normalize the expression by $\sigma_0$, the partonic Born cross section. The interested reader can find explicit expressions for $\cN_{ij}$ and $\df\Phi_{h+n}$ in \refcite{Ebert:2020lxs}.

In \refcite{Ebert:2020lxs}, we showed that the matching coefficient in \eq{beam_OPE} is obtained by taking the limit of \eq{sigma_part} where all real and loop momenta are treated as being collinear to $n$-direction, which is referred as the \emph{strict} $n$-collinear limit, and we refer to \refcite{Ebert:2020lxs} for details on its calculation:
\begin{align} \label{eq:Iij_naive}
 \cI_{ij}^{\rm naive}(z,q_T)
 = \int_0^1 \df x \int_0^\infty \df \wa\df \wb \,&
   \delta[z - (1-\wa)] \, \delta\Big[q_T^2 - \frac{\wa\wb}{1-\wa} (1-x) Q^2\Bigr]
   \nn\\&\times\,
   \nlim \frac{\df\eta_{ij}}{\df Q^2  \df\wa \df\wb \df x}
\,.\end{align}
Solving the $\delta$ functions fixes $\wa$ and $\wb$ as
\begin{align} \label{eq:vars_LP}
 \wa = 1-z \,,\quad
 \wb = \frac{q_T^2}{Q^2} \frac{z}{(1-x)(1-z)}
\,.\end{align}
The superscript $^{\rm naive}$ in \eq{Iij_naive} indicates that further steps are required to obtain the desired matching kernel. First, we note that the integral in \eq{Iij_naive} contains the aforementioned rapidity divergences, namely divergences as $x\to1$ or $z\to1$ that are not regulated by dimensional regularization. We regulate these using the exponential regulator of \refcite{Li:2016axz}, which in fact is the only regulator in the literature compatible with our approach.
Inserting the regulator factor $\exp(2 \tau e^{-\gamma_E} k^0)$ expressed in the above variables, we obtain the regulated kernel as \cite{Ebert:2020lxs}
\begin{align} \label{eq:Iij_bare}
 \cI_{ij}(z,q_T,\eps,\tau/\omega) = \lim_{\substack{\tau\to0\\\eps\to0}} \int_0^1\df x &\,\frac{z}{Q^2 (1-x)(1-z)}
 \exp\left[ - \tau e^{-\gamma_E} \frac{q_T^2}{\omega} \frac{z}{(1-x)(1-z)}\right]
 \nn\\&\times
 \nlim \frac{\df\eta_{ij}}{\df Q^2  \df \wa \df \wb \df x}\biggl|_{\eqref{eq:vars_LP}}
\,.\end{align}
Here, $\omega = Q e^Y$ is the so-called label momentum. The exponential factor in \eq{Iij_bare} regulates divergences as $x,z\to1$, with $\tau$ being the rapidity regulator. UV and IR divergences are regulated by working in $d=4-2\eps$ dimensions, with the limit $\eps\to0$ being taken after the limit $\tau\to0$, as indicated. It is convenient to expand the exponential factor in \eq{Iij_bare} in terms of distributions before carrying out the integral~\cite{Luo:2019hmp}, and we provide more details on this in \app{rapidity_regulator}.

It is common to express TMD beam functions in Fourier space, where convolutions in $\qt$ are replaced by simple products, which in particular greatly simplifies the resummation of large logarithms~\cite{Ebert:2016gcn}.
Denoting the Fourier-transform matching kernel as $\tilde\cI_{ij}$, we obtain the renormalized kernel as
\begin{equation} \label{eq:Iij_ren}
 \tilde \cI_{ij}(z,b_T,\mu,\nu/\omega)
 = \sum_{k} \int_z^1\!\frac{\df z'}{z'} \, \Gamma_{jk}\Bigl(\frac{z}{z'}, \eps\Bigr) \,
   \tilde Z_B^i(\eps,\mu,\nu/\omega) \, \hat Z_{\as}(\mu,\eps) \,
   \frac{\tilde \cI_{i k}(z',b_T,\eps,\tau)}{\tilde S(b_T,\eps,\tau)}
.\end{equation}
Here, following \refcite{Li:2016ctv} we identify $\nu \equiv 1/\tau$ as the rapidity renormalization scale~\cite{Chiu:2012ir}.
The so-called zero-bin subtraction~\cite{Manohar:2006nz} to subtract overlap of the beam function with the soft function is implemented by dividing by the soft function $\tilde S$~\cite{Luo:2019hmp}.
In \eq{Iij_ren}, the counterterm $\hat Z_{\as}$ implements the renormalization of the bare coupling constant $\as^b$ in the $\MSbar$ scheme as stated in \eq{asb}.
Infrared divergences are canceled through the convolution with the PDF counterterm $\Gamma_{jk}$, which is given in \eq{Gamma}.
The remaining poles in $\eps$ are of UV nature in SCET and are thus canceled by an additional UV counter term in the effective theory, which is the beam function counter term $\tilde Z_B$.

Since the bare soft function is not given in the literature, we have directly calculated it from the soft limit of \eq{sigma_part} similar to \eq{Iij_bare}.
\begin{align} \label{eq:S_bare}
 S (q_T,\eps,\tau)
 = \lim_{\substack{\tau\to0\\\eps\to0}}&
   \int_0^1\df x \int_0^\infty \, \df w_1 \df w_2 \,
   \delta\bigl[q_T^2 - \wa\wb (1-x) Q^2\bigr]
\nn\\&\times
   \exp\bigl[-2 Q \tau e^{-\gamma_E} (\wa + \wb)\bigr]
   \lim\limits_{\text{strict soft}} \frac{\df\eta_{ij}}{\df Q^2  \df \wa \df \wb \df x}
\,.\end{align}
In the strict soft limit, both $\wa$ and $\wb$ are treated as small quantities, such the measurement $\delta$ function and the exponential regulator in \eq{S_bare} are simpler than in \eq{Iij_naive}.
The Fourier transform of \eq{S_bare} yields the bare soft function $\tilde S(b_T,\eps,\tau)$ required in \eq{Iij_ren}. We have also verified that the renormalized soft function reproduces the N$^3$LO result of \refcite{Li:2016ctv}.
Since the bare soft function in the exponential regulator is only given at NLO in the literature~\cite{Li:2016axz}, we provide the bare soft function in electronic format in the ancillary files.

The strict soft limit of the general differential partonic coefficient function is obtained by expanding the strict collinear limit in $w_1$ and maintaining only the first term of the generalised power series.
At $n^{th}$ order in perturbation theory  the strict soft limit of the partonic coefficient function takes the form
\beq \label{eq:soft_limit}
\lim\limits_{\text{strict soft}} \frac{\df\eta_{ij}^{(n)}}{\df Q^2  \df \wa \df \wb \df x}= \omega_1^{-1-n\eps} \omega_2^{-1-n\eps} \eta^{(n)}_{\text{strict soft}} (x,\eps),
\eeq
where $\eta^{(n)}_{\text{strict soft}} (x,\eps)$ is independent of $w_1$ and $w_2$. 
Note that \eq{soft_limit} is related to the bare fully-differential soft function which measures the total soft radiation in a process.
This limit is also easily related to the bare threshold soft function~\cite{Li:2014afw} via
\beq
S_{\text{thr}}^{(n)}(z,\epsilon)=\int_0^\infty \df w_1 \, \df w_2 \int_0^1 \df x \, \delta(1-z-w_1-w_2 ) \eta^{(n)}_{\text{strict soft}} (x,\eps)\,,
\eeq
where $z$ is the threshold parameter.
We have used this relation as an additional check on our soft limit.

\section{Results}
\label{sec:results}

Here, we present our results for the matching kernels of the TMD beam functions at N$^3$LO.
Our calculation leverages on the collinear expansion of the cross sections for off-shell photon production (Drell-Yan) and Higgs production in gluon fusion in proton-proton collisions.
The computation of the Higgs boson production cross section is performed in the heavy top quark effective theory where the gluons are directly coupled to the Higgs boson via an effective operator generated by integrating out the top quark field from the SM Lagrangian~\cite{Inami1983,Shifman1978,Spiridonov:1988md,Wilczek1977,Chetyrkin:1997un,Schroder:2005hy,Chetyrkin:2005ia,Kramer:1996iq,Kniehl:2006bg}. 
The matrix elements for this computation can be categorized by the number of final state partons in addition to the color singlet final state. The relevant matrix elements for the calculation of the N$^3$LO differential cross sections in the collinear limit involve one (RVV), two (RRV) and three (RRR) final state partons.

The results for the partonic cross sections involving matrix elements with exactly one parton in the final state are available in full kinematics from refs.~\cite{Duhr:2020seh,Dulat:2014mda,Dulat:2017brz,Dulat:2017prg} which build on previous work done in refs.~\cite{Anastasiou:2013mca,Duhr:2014nda,Duhr:2013msa}.
Therefore, in order to obtain the RVV contributions in the strict collinear limit, we can straightforwardly expand the results in full kinematics and extract the required components.

To compute the collinear limit of the partonic cross sections with more than one final state parton, the necessary Feynman diagrams are obtained using QGRAF~\cite{Nogueira_1993}. We carry out the spinor and color algebra using an in-house code, and perform the strict collinear expansion of these matrix elements following the procedure outlined in ref.~\cite{Ebert:2020lxs}.
In order to integrate over loop and phase space momenta with measurement and on-shell constraints, we make use of the framework of reverse unitarity~\cite{Anastasiou2003,Anastasiou:2002qz,Anastasiou:2003yy,Anastasiou2005,Anastasiou2004a}.

We re-express our expanded cross section in terms of master integrals (MI) via integration-by-parts (IBP) identities~\cite{Chetyrkin:1981qh,Tkachov:1981wb}. We obtain a basis of 492 MI, expressed in terms of the variables in \eq{vardef} as well as the dimensional regularization parameter $\epsilon$, of which 172 MI are required to describe the RRV contributions, while 320 are needed for the RRR ones.
In order to compute the collinear master integrals we employ the method of differential equations~\cite{Kotikov:1990kg,Kotikov:1991hm,Kotikov:1991pm,Henn:2013pwa,Gehrmann:1999as}.
We fix the boundary conditions for the differential equations by further expanding the collinear master integrals in the soft limit and integrating over the phase space, such that the result of this procedure can then be easily matched to the soft integrals calculated in \refscite{Anastasiou:2013srw,Anastasiou:2014lda,Anastasiou:2014vaa,Anastasiou:2015yha,Duhr:2019kwi}.

Completing these steps we obtain the bare differential partonic cross section expanded in the strict collinear limit of \eq{sigma_part}.
We note that the ingredients computed so far are the same as those neeeded for the calculation of the N$^3$LO $N-$jettiness beam functions of \refcite{Ebert:2020unb}.
Next, we obtain the bare matching kernel via \eq{Iij_bare} and subsequently perform the Fourier transform over $\qt$ using \eq{FT_qT}.
Both the calculation of the differential partonic cross section as well as the extration of the $\qt$-dependent matching kernels will be elaborated in ref.~\cite{Ebert:ThingsToCome}.
Finally, the renormalized matching kernel is obtained using \eq{Iij_ren},
where the beam function counter term $\tilde Z_B^i$ was predicted from the renormalization group equations (RGEs) of the beam function as shown in \app{counter_term}.
The TMDPDF can then be straightforwardly obtained by combining the beam function with the soft function as in \eq{TMDPDFdef}.

We expand the matching kernels $\tilde\cI_{ij}$ of the beam function and the matching kernels $\tilde\cI^{\rm TMD}_{ij}$ of the TMDPDF obtained in this way in powers of $\as/\pi$,
\begin{align}
 \tilde\cI_{ij}(z,b_T,\mu,\nu/\omega) &
 = \sum_{\ell=0}^\infty \Bigl(\frac{\as}{\pi}\Bigr)^\ell \tilde\cI_{ij}^{(\ell)}(z,b_T,\mu,\nu/\omega)
\,,\nn\\
 \tilde\cI^{\rm TMD}_{ij} (z,b_T,\mu,\zeta) &
 = \sum_{\ell=0}^\infty \Bigl(\frac{\as}{\pi}\Bigr)^\ell \tilde\cI^{\mathrm{TMD}\,(\ell)}_{ij}(z,b_T,\mu,\zeta)
\,,\end{align}
where the coefficients $\tilde\cI_{ij}^{(\ell)}$ and $\tilde\cI^{\mathrm{TMD}~(\ell)}_{ij}$ can be written as a polynomial in logarithms of the appearing scales with $z$-dependent coefficient functions,
\begin{align}\label{eq:logdef}
 \tilde\cI_{ij}^{(\ell)}(z,b_T,\mu,\nu/\omega) &= \sum_{m,n=0}^\ell \tilde\cI_{ij}^{(\ell,m,n)}(z) L_b^n L_\omega^m
\,,\\\nn
 \tilde\cI^{\mathrm{TMD}\,(\ell)}_{ij}(z,b_T,\mu,\zeta) &= \sum_{n=0}^{2\ell} \sum_{m=0}^{\ell} \tilde\cI^{\mathrm{TMD}~(\ell,m,n)}_{ij}(z) L_b^n L_\zeta^m
\,.\end{align}
The logarithms in \eq{logdef} are defined as
\begin{align}
 L_b = \ln\frac{b_T^2 \mu^2}{4 e^{-2\gamma_E}}
 \,,\qquad L_\omega = \ln\frac{\nu}{\omega}
 \,,\qquad L_\zeta = \ln\frac{\mu^2}{\zeta}
\,,\end{align}
where $\gamma_E$ is the Euler-Mascheroni constant, and we remind the reader that $\zeta=\omega^2$ is the energy of the struck parton.
The logarithmic terms with $m > 0$ or $n > 0$ in \eq{logdef} fully describe the scale dependence of both the TMDPDF as well as of the beam function. Therefore, their structure is completely determined by the beam function RGEs (see \app{RGEs}) in terms of its anomalous dimensions and lower-order ingredients.
The nonlogarithmic beam function boundary term at N$^3$LO
\begin{equation}
 \tilde I_{ij}^{(3)}(z) \equiv \tilde\cI_{ij}^{(3,0,0)}(z)
\,,\end{equation}
is the genuinely new result calculated by us in this work.
Remarkably, it can be expressed entirely in terms of standard plus distributions and harmonic polylogarithms~\cite{Remiddi:1999ew} of argument $z$ and transcendental weight less or equal to five.
To allow for an easy numeric implementation, we also provide a generalized power series expansion of our results around $z=0$ and $z=1$ with up to 50 terms in each expansion.
Both expansions formally converge in the unit interval, but, clearly, the convergence of the series improves as the expansion parameter gets smaller.
In the ancillary files of the arXiv submission of this article we provide both power series as well as the analytic solution for all matching kernels.

We performed several checks on our results. Firstly, we verified that the UV and IR subtraction as given in \eq{Iij_ren} correctly removes all poles in $\eps$. Our NLO and NNLO results for the renormalized beam function are validated against \refcite{Luebbert:2016itl}, and the bare results through $\cO(\eps^4)$ at NLO and through $\cO(\eps^2)$ are verified against \refscite{Luo:2019hmp,Luo:2019bmw}.
Given that the logarithmic terms in \eq{logdef} are dictaded by the beam function RGE, we verify that all logarithmic terms of our result are correct by comparing them against those predicted in \refcite{Billis:2019vxg} by solving the beam function RGE.
We also verified the eikonal limit
\begin{align} \label{eq:eikonal_limit}
 \lim_{z\to1} \tilde I_{ij}^{(3)}(z) &
 = \frac{\gamma_2^r}{64} \, \delta_{ij} \, \cL_0(1-z)
\,,\end{align}
which was derived in \refcite{Billis:2019vxg} from consistency with joint $q_T$ and soft threshold resummation relations~\cite{Li:2016axz,Lustermans:2016nvk}, and also conjectured in \refcite{Echevarria:2016scs}.
In \eq{eikonal_limit}, $\gamma_2^r$ is the three-loop coefficient of the so-called rapidity anomalous dimension~\cite{Chiu:2012ir}, as given in Eq.~(16) of \refcite{Li:2016ctv} (see also \refcite{Vladimirov:2016dll}), where the appropriate color structure is implicit.
The rapidity anomalous dimensions is also closely related to the Collins-Soper kernel of \refscite{Collins:1981uk,Collins:1981va}.
For the quark beam function, we also compared with the results recently obtained in \refcite{Luo:2019szz}. We find discrepancies for terms proportional to the color structure $d^{abc}d_{abc}$ entering in all quark-to-quark kernels. After private communication, the authors of~\refcite{Luo:2019szz} identified and resolved a minor mistake in their calculation, after which they find agreement with our result.
Furthermore, another check of our results comes from the fact that the first four terms in the soft expansion of the Higgs cross section correctly match the collinear limit of the threshold expansion of the partonic cross section obtained in ~\refcite{Dulat:2017prg,Dulat:2018bfe}.
We also note that inclusive cross section for Drell-Yan as well as for Higgs production at N$^3$LO was calculated in \refscite{Anastasiou:2015ema,Duhr:2020seh,Mistlberger:2018etf,Anastasiou:2014lda,Anastasiou:2014vaa}.
Using the collinear partonic coefficient functions of our calculation after integration over phase space, we also correctly reproduce the leading threshold expansion contribution of all partonic initial states that contribute to the collinear limit of the partonic cross sections.

\begin{figure*}
 \centering
 \includegraphics[width=0.49\textwidth]{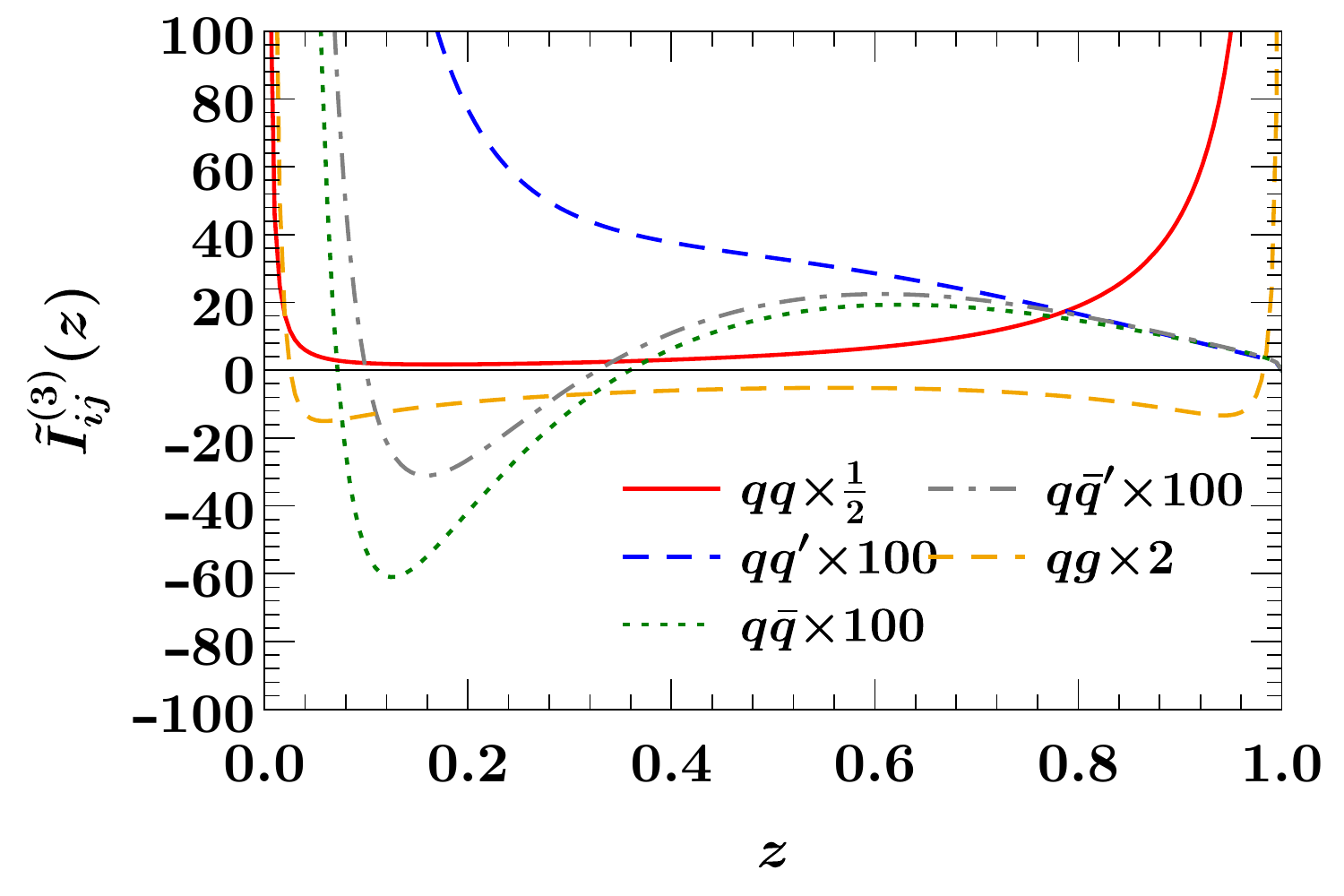}
 \hfill
 \includegraphics[width=0.49\textwidth]{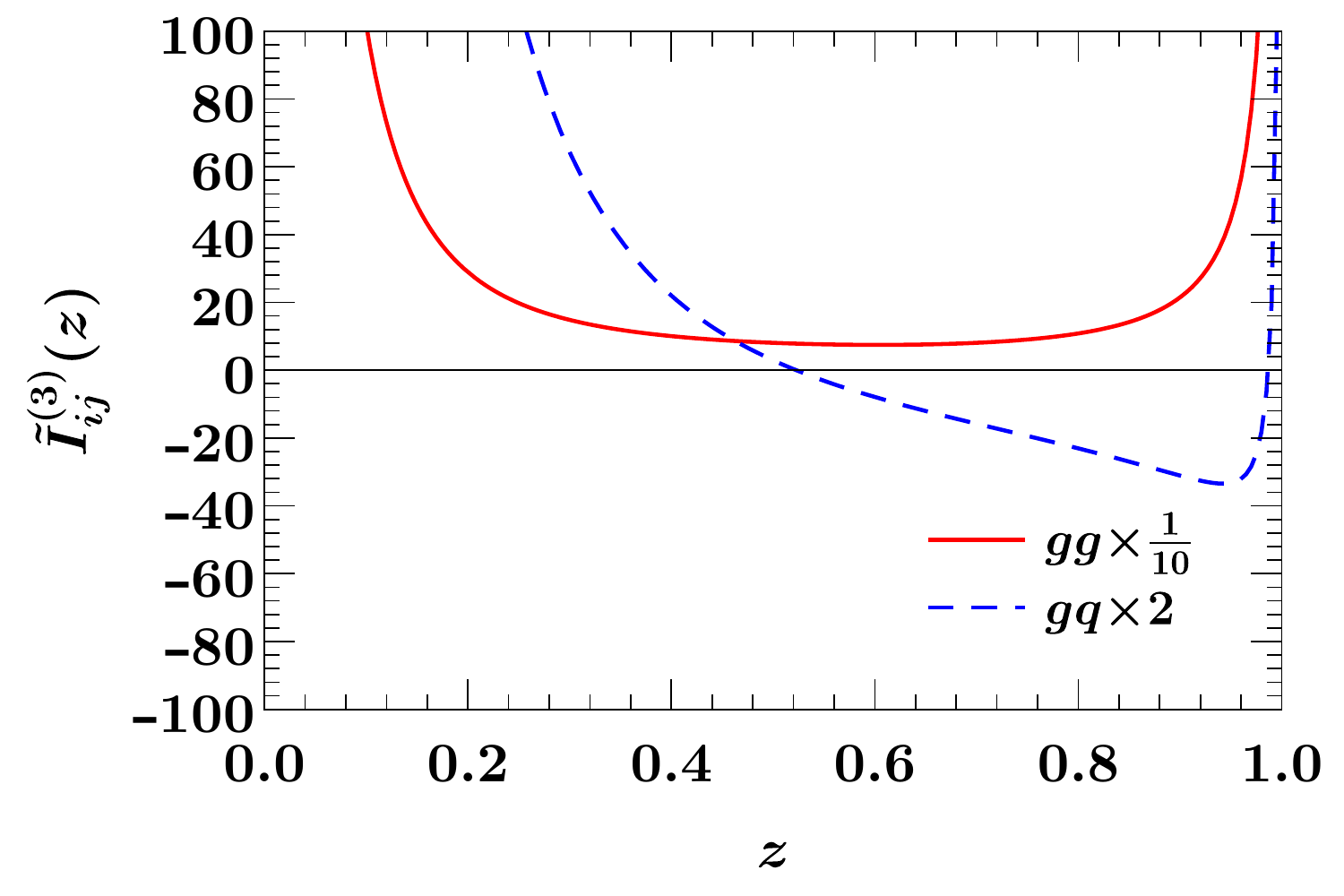}
 \caption{The N$^3$LO beam function boundary term $\tilde I^{(3)}_{ij}(z)$ as a function of $z$ in all channels contributing to the quark beam function (left) and the gluon beam function (right). The different channels are rescaled as indicated in the figures.}
 \label{fig:I3}
\end{figure*}

Let us numerically illustrate our results. In \fig{I3} we plot the beam function boundary terms $\tilde I_{ij}(z)$ for the quark (left) and gluon (right) beam functions as a function of $z$. Note that in this plot we set $\delta(1-z) \to 0$ and replaced the distribution $\cL_0(1-z) \to (1-z)^{-1}$.
For illustration purposes we rescaled the different channel as indicated, given that they give rise to very different shapes and magnitudes.

Next, we study the impact of our calculation on the beam function and TMDPDF themselves.
We use the \texttt{MMHT2014nnlo68cl} PDF of \refcite{Harland-Lang:2014zoa}, using $\as(m_Z)=0.118$ and the evaluation of \eq{beam_OPE} is obtained through an implementation of our results in \texttt{SCETlib}~\cite{scetlib}.

\begin{figure*}
 \centering
 \includegraphics[width=0.49\textwidth]{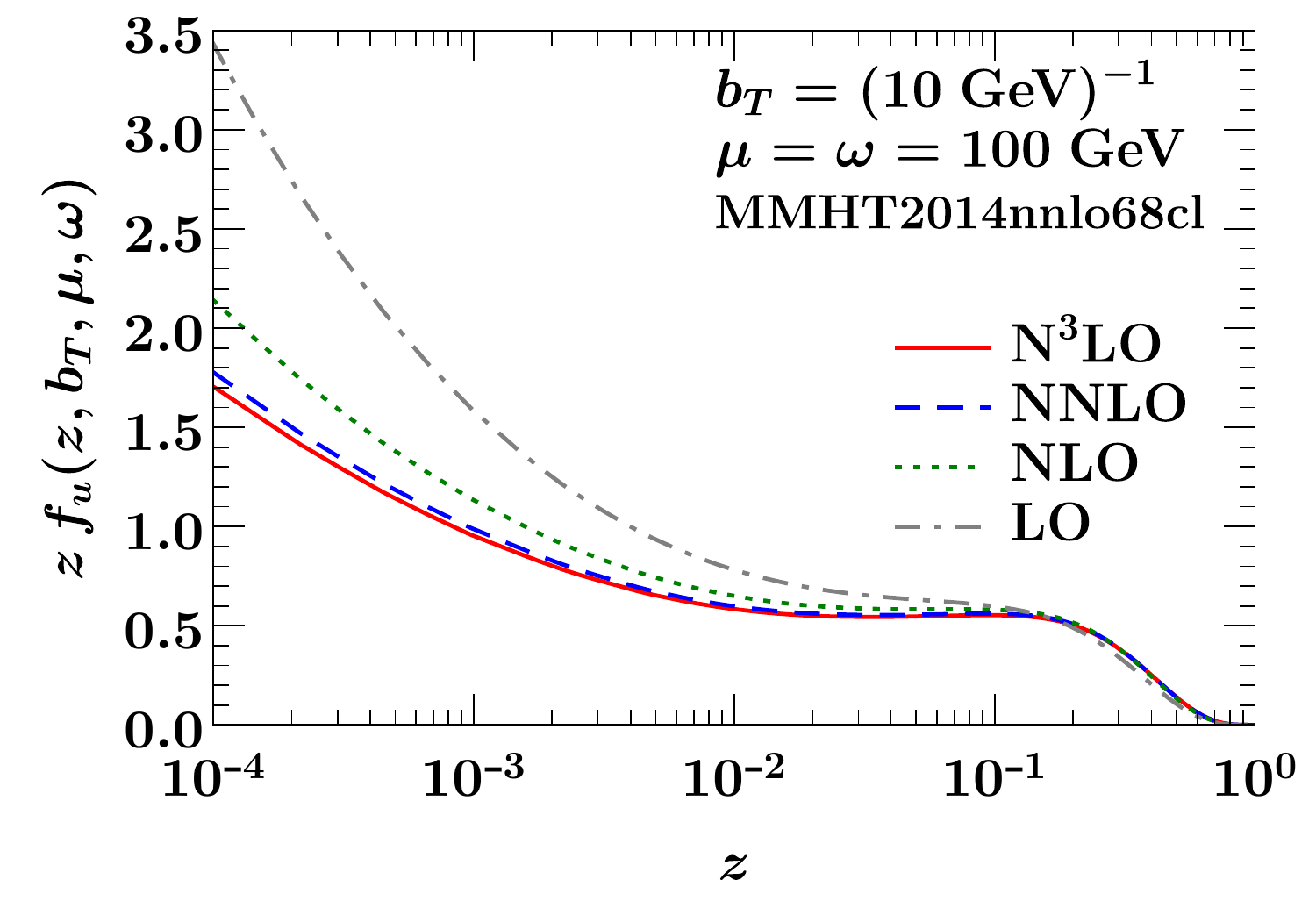}
 \hfill
 \includegraphics[width=0.49\textwidth]{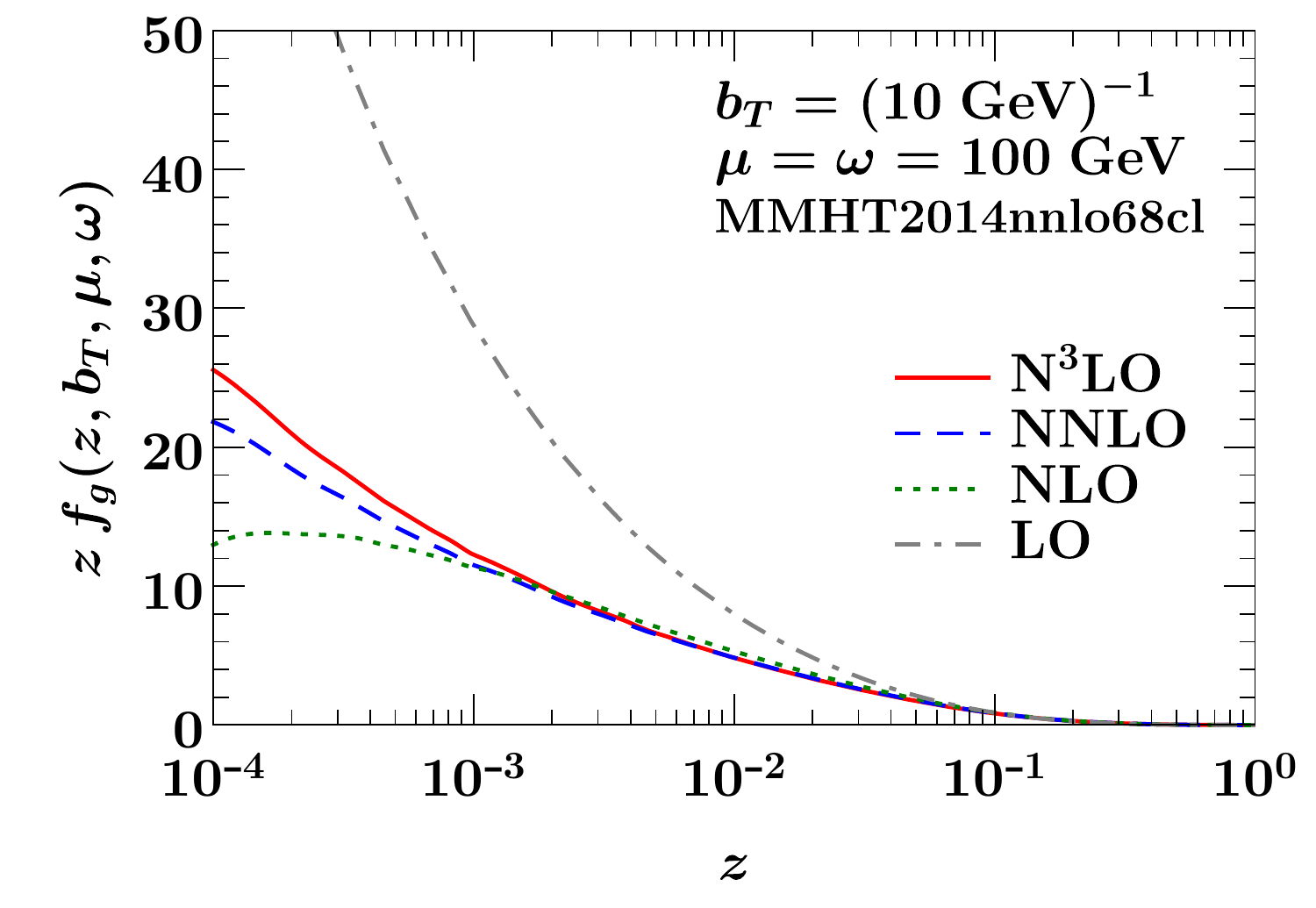}
 \caption{The $u$-quark TMDPDF (left) and the gluon TMDPDF (right) as a function of $z$ for fixed $b_T = (10~\GeV)^{-1}$ and $\mu=\omega=100~\GeV$. We show the result at LO (which corresponds to the PDF), NLO, NNLO and N$^3$LO.}
 \label{fig:TMDPDF_convergence}
\end{figure*}

In \fig{TMDPDF_convergence}, we show the $u$-quark TMDPDF (left) and the gluon TMDPDF (right) at different orders in the coupling constant%
\footnote{Note that while varying the perturbative order of the matching kernel we keep the \texttt{MMHT2014nnlo68cl} PDF fixed. It is also interesting to study the simultaneously variation of both the order of the matching kernel as well as that of the PDF on which the beam function gets matched onto. However, an extraction of PDFs at N$^3$LO is currently not available and while there are methods to the study the uncertainty due to missing higher order PDFs \cite{Anastasiou:2016cez,Forte:2013mda}, this is clearly independent of the calculation of the matching kernel.}
as a function of $z$.
We fix the impact parameter $b_T = 10~\GeV^{-1}$, parton energy $\omega = 100~\GeV$ and renormalization scale $\mu = 100~\GeV$.
Since the LO result for the beam function corresponds to the PDF itself, \fig{TMDPDF_convergence} can be used to appreciate the difference in shape of the beam function compared to the PDF. With the inclusion of the N$^3$LO result obtain in this work, both the quark and the gluon TMDPDFs nicely show convergence over a large range of values for $z$.

\begin{figure*}
 \centering
 \includegraphics[width=0.49\textwidth]{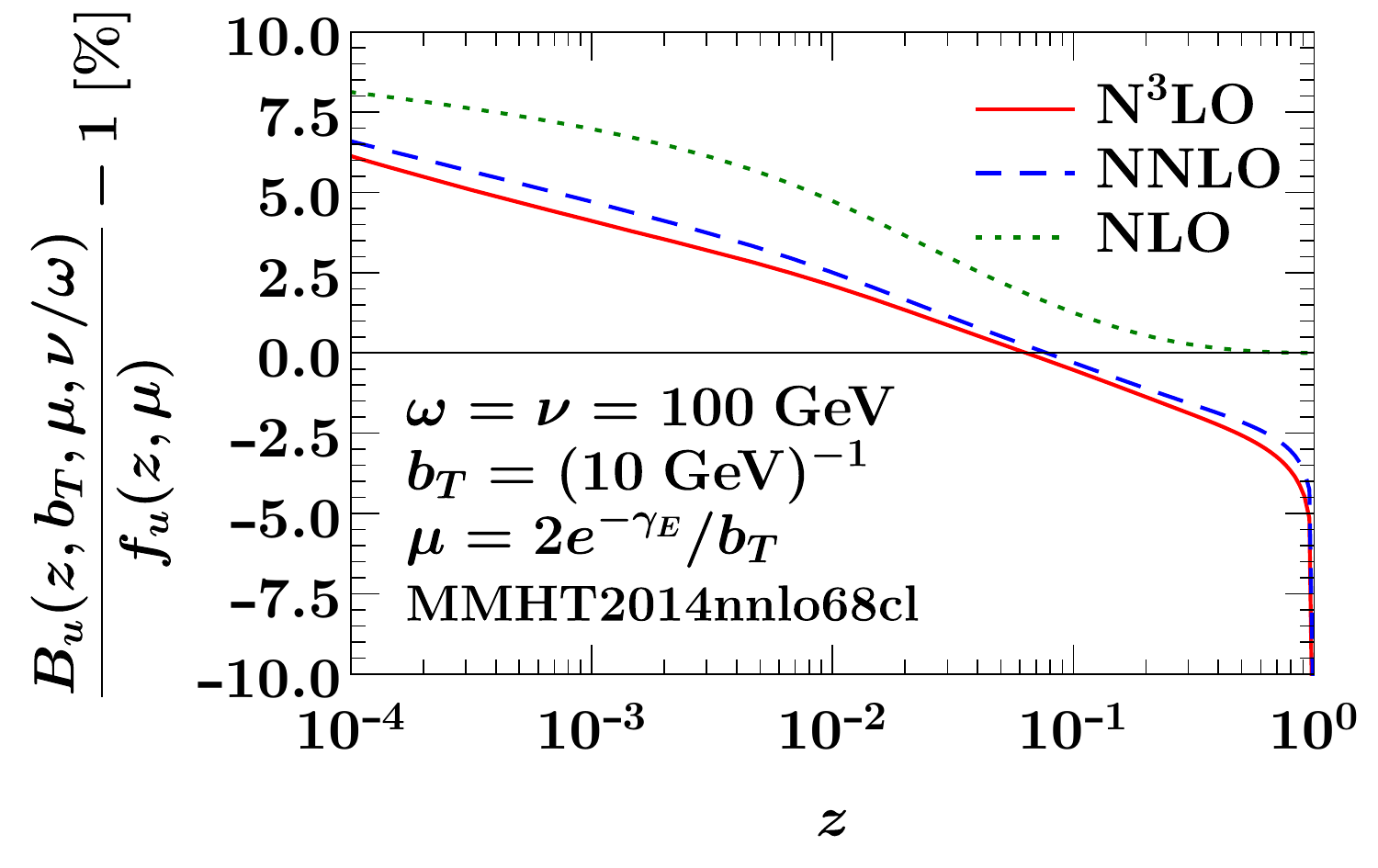}
 \hfill
 \includegraphics[width=0.49\textwidth]{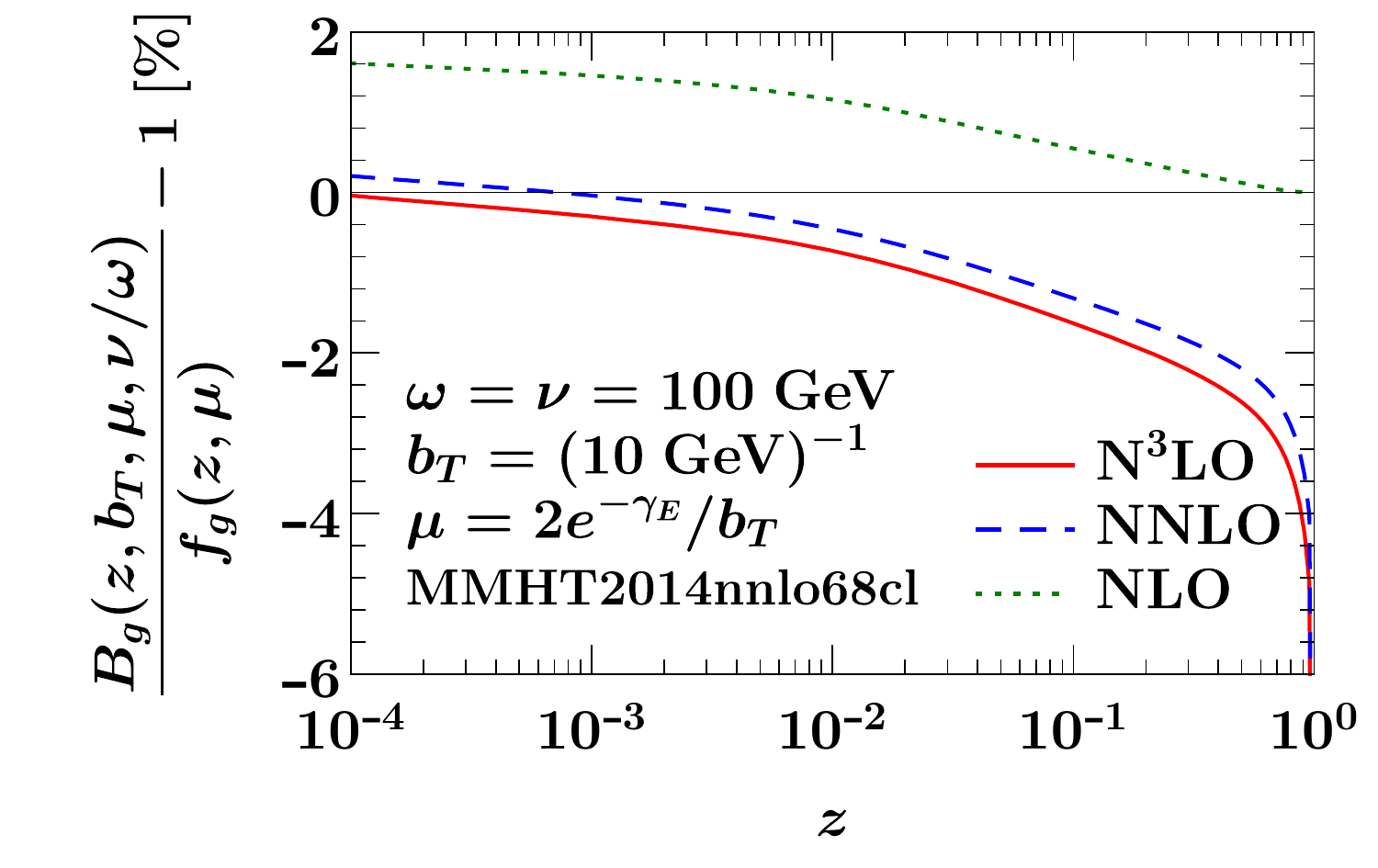}
 \caption{The relative difference of $u$-quark beam function (left) and the gluon beam function (right) to the corresponding PDF, as a function. We fix $b_T = (10~\GeV)^{-1}$ and $\omega=100~\GeV$ and choose the canonical scales $\mu b_T = 2 e^{-\gamma_E}$ and $\nu=\omega$. Note that with this choice of scales, the displayed beam function is the boundary term of a resummed prediction.}
 \label{fig:ratio_to_LO}
\end{figure*}

In order to understand the impact of the new three-loop boundary term $\tilde I_{ij}^{(3)}$ in a resummed predictions, we present the beam function evaluated at the canonical scales $\mu b_T = 2 e^{-\gamma_E}$ and $\nu = \omega$, where all logarithms in \eq{logdef} vanish and only the boundary term $\tilde I_{ij}^{(3)}$ contributes.
In \fig{ratio_to_LO}, we compare the $u$-quark beam function (left) and gluon beam function (right) order by order in $\alpha_s$, up to N$^3$LO, to the corresponding PDFs, choosing canonical scales for $b_T = (10~\GeV)^{-1}$ and $\omega=100~\GeV$.
We see that the beam function has a very different shape compared to the PDF, and that the beam function converges very well at N$^3$LO.

\begin{figure*}
 \centering
 \includegraphics[width=0.55\textwidth]{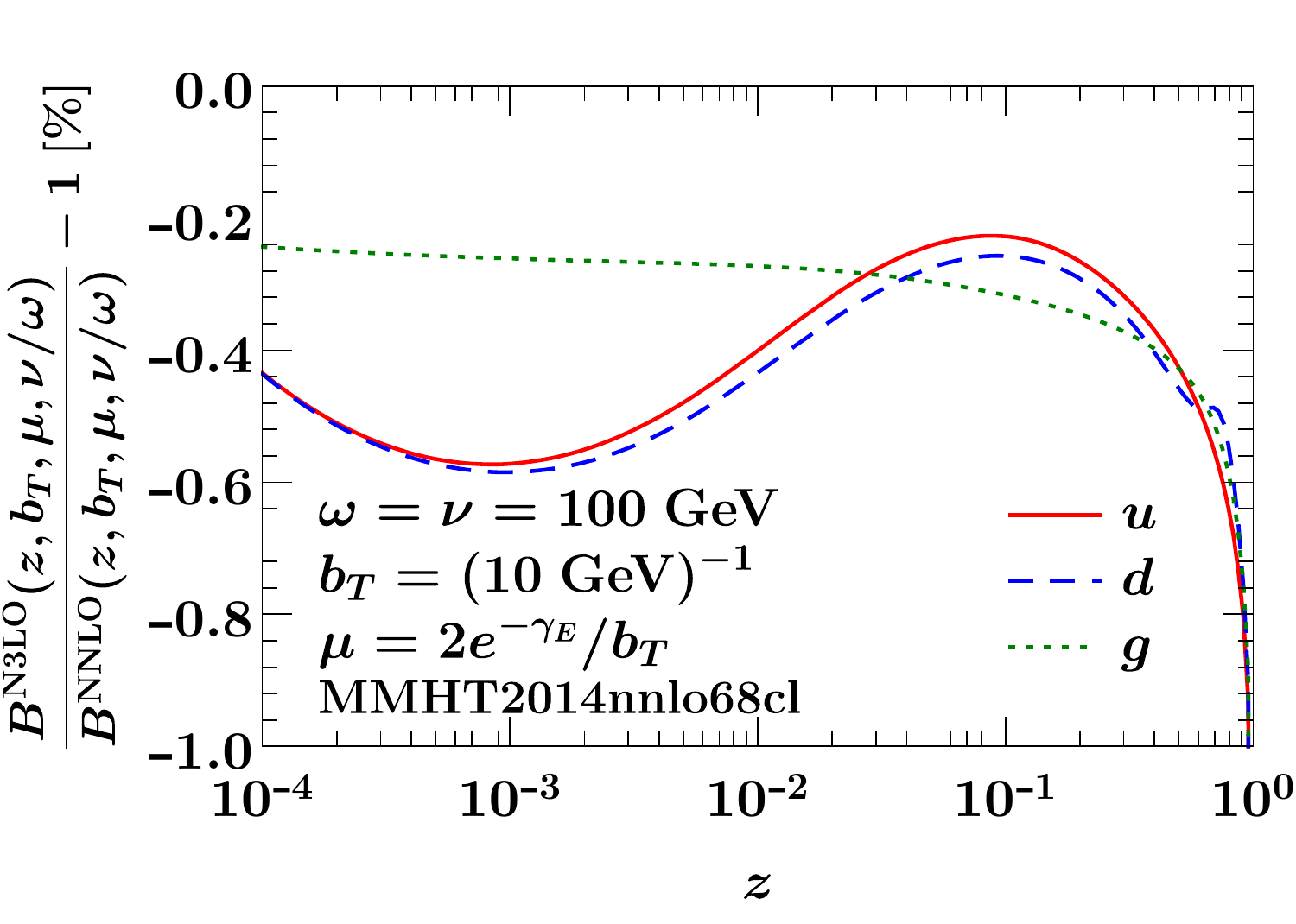}
 \caption{The $K$-factor of the N$^3$LO beam function, i.e.\ the ratio of the N$^3$LO beam function to NNLO beam function. We fix $b_T = (10~\GeV)^{-1}$ and $\omega=100~\GeV$ and choose the canonical scales $\mu b_T = 2 e^{-\gamma_E}$ and $\nu=\omega$, such that the shown beam function corresponds to the boundary term in a resummed prediction. The different colors show the results for an $u$-quark, $d$-quark and gluon, respectively.}
 \label{fig:Kfactor}
\end{figure*}

Finally, the $K$-factor of the N$^3$LO beam function, which is defined as the ratio of the beam function at N$^3$LO w.r.t. its value at NNLO, is shown in \fig{Kfactor}.
As before, we choose the canonical scales for $b_T = (10~\GeV)^{-1}$ and $\omega=100~\GeV$.
We find a rather small correction of $\sim 0.2-0.5\%$, but with a notable dependence on $z$ for all channels.

For completeness, we also present the high-energy limit $z\to0$ of the kernels $\tilde I_{gg}^{(3)}(z)$ and $\tilde I_{gq}^{(3)}(z)$ contributing to the gluon beam function in \app{high_energy}. The corresponding limit for the quark kernels were already presented in \refcite{Luo:2019szz}, for which we find perfect agreement.
These results are useful to study the small-$x$ behavior of TMDPDFs, see e.g.~\refscite{Balitsky:2015qba,Marzani:2015oyb,Balitsky:2016dgz,Xiao:2017yya}.

\section{Conclusions}
\label{sec:conclusion}

We have calculated the perturbative matching kernel relating transverse-momentum dependent beam functions with lightcone PDFs at N$^3$LO in QCD.
This provides the first results of these kernels for the gluon TMD beam function, and corrects the result in the $d_{abc} d^{abc}$ color structure in the recent calculation of the quark TMD beam function in \refcite{Luo:2019szz}.
After private communication, the authors of~\refcite{Luo:2019szz} identified and resolved a minor mistake in their calculation, after which perfect agreement is found.
This emphasizes that having two independent computations that are in accordance with each other is the most stringent cross check.
Our results are obtained via a framework recently developed by us that allows for the efficient expansion of differential hadronic collinear cross sections~\cite{Ebert:2020lxs}, showing its applicability for the extraction of universal ingredients arising in the collinear limit of QCD to an unprecedented level of precision in perturbation theory.
As a byproduct, we also confirmed a previous computation of the soft function for transverse momentum factorization~\cite{Li:2016axz}.

The results of our calculation are provided in the ancillary files of this paper submission. We include, for all quark and gluon channels, the renormalized TMD beam functions, along with their expansions around $z=0$ and $z=1$ up to 50 orders in each expansion, as well as the renormalized TMDPDFs and the bare and renormalized soft function.

The phenomenological applications of our results are numerous.
Firstly, we provide the last missing ingredient for the fully-differential calculation of color-singlet processes at N$^3$LO using the $q_T$ subtraction method~\cite{Catani:2007vq,Cieri:2018oms,Billis:2019vxg}, which can be used to obtain the first exact fully-differential cross sections at this order.
They also enable the resummation of transverse momentum distributions at hadron colliders at N$^3$LL$^\prime$ accuracy, both for gluon and quark induced hard scatterings such as Higgs-boson production and the Drell-Yan process, two key observables at the LHC.

A natural future direction of this research is the calculation of the closely related TMD fragmentation functions at N$^3$LO, which are required to describe the small-$q_T$ limit semi-inclusive deep inelastic scattering and are currently only known at NNLO \cite{Echevarria:2016scs,Luo:2019bmw,Luo:2019hmp}.

With the same techniques presented in this work, it would also be interesting to consider the calculation of other $\qt$-dependent time-like collinear functions, such as the Energy-Energy Correlator (EEC) jet functions.
The quark and gluon EEC jet functions enter the factorization of the EEC in the back-to-back limit for $e^+e^-$ annihilation and gluon initiated Higgs decay, respectively~\cite{Moult:2018jzp}, as well as that for the TEEC~\cite{Gao:2019ojf}.
Their knowledge at N$^3$LO is the only missing ingredient to achieve resummation of the EEC in the back-to-back limit at N$^3$LL$^\prime$ accuracy. Note that at fixed order the full angle EEC has been calculated analytically through $\cO(\alpha_s^2)$ \cite{Dixon:2018qgp,Luo:2019nig}.

It will also be interesting to study the collinear expansions beyond leading power to shed light on the structure of TMD factorization at subleading power, in particular on the structure of rapidity divergences at subleading power~\cite{Moult:2017xpp,Ebert:2018gsn,Moult:2019vou}.

\acknowledgments
We thank the Ming-xing Luo, Tong-Zhi Yang, Hua Xing Zhu and Yu Jiao Zhu for private communication concerning the results of \refcite{Luo:2019szz} prior to the publication of this article that allowed both groups to agree on our results.
We also thank Johannes Michel, Iain Stewart and Frank Tackmann for useful discussions.
This work was supported by the Office of Nuclear Physics of the U.S.\ Department of Energy under Contract No.\ DE-SC0011090 and DE-AC02-76SF00515, and within the framework of the TMD Topical Collaboration.
M.E.\ is also supported by the Alexander von Humboldt Foundation through a Feodor Lynen Research Fellowship,
and B.M.\ is also supported by a Pappalardo fellowship.

\appendix

\section{Ingredients for the calculation of the beam function}
\label{app:details}

In this appendix, we provide more details on the regularization and renormalization of the beam function kernels. Details of the calculation of all required integrals will be presented in \refcite{Ebert:ThingsToCome}.

\subsection{Rapidity regularization}
\label{app:rapidity_regulator}

In practice, it is useful to expand the regulator in \eq{Iij_bare} in terms of plus distributions, which allows one to take the limit $\tau\to0$ before evaluating the integral in \eq{Iij_bare}, see also \cite{Luo:2019hmp} for a discussion at two loops.
In general we want to expand a power divergence using the relation
\beq
\label{eq:testfuncexp}
\lim\limits_{\tau\to 0} \int_0^1\df x \, e^{-\tau/x} \frac{f(x)}{x^{1-a}} = \int_0^1\df x \left( f(0) \,\delta(x) \lim\limits_{\tau\to 0} G(\tau) + f(x)\left[\frac{1}{x^{1-a}} \right]_+ \right).
\eeq
Here, $f(x)$ is a suitable test function that is holomorphic at $x=0$ and $a$ represents a generalized power, typically proportional to the dimensional regulator $\eps$.
We find that
\beq
G(\tau)=\int_0^1\df x\, \frac{e^{-\tau/x}}{x^{1-a}} =\frac{1}{a}\bigl[1 - \tau^a \Gamma(1-a)\bigr] + \cO(\tau).\\
\eeq
We furthermore require the two-dimensional generalization of the above integral,
\bea
\int_0^1\!\df x\,\df y \,  e^{-\tau / (x y)} x^{-1+a} y^{-1+b}
&=& \frac{1}{ab} - \frac{\tau^a \Gamma(-a)}{a-b} + \frac{\tau^b \Gamma(-b)}{a-b} + \cO(\tau)
\eea
The regularization of an integral of the type of the above equation including a test function follows along the lines of \eq{testfuncexp}.

In order to compute the soft function we use the following relations
\begin{align} \label{eq:soft_regulator}
 &\lim_{\tau\to 0} \int_0^\infty \df z_1 \, \df z_2 \int_0^1 \df\xb \,
   (z_1 z_2 \xb)^{-1+a\eps} \, \delta\bigl(q_T^2 - \xb z_1 z_2\bigr) e^{-2\tau e^{-\gamma_E}(z_1+z_2)}
\nn\\&\quad
 = (q_T^2)^{-1+a\eps} \biggl[ \frac12 \ln^2(4\tau^2 q_T^2) + \frac{\pi^2}{6}  \biggr]
\,,\\
 &\lim_{\tau\to 0} \int_0^\infty \df z_1 \, \df z_2 \int_0^1 \df\xb \,
   (z_1 z_2 \xb)^{-1+a\eps} \, \delta\bigl(q_T^2 - \xb z_1 z_2\bigr) e^{-2\tau e^{-\gamma_E}(z_1+z_2)} \ln\xb
\nn\\&\quad
 = (q_T^2)^{-1+a\eps} \biggl[ \frac16 \ln^3(4\tau^2 q_T^2) + \frac{\pi^2}{6} \ln(4\tau^2 q_T^2) + \frac23 \zeta_3 \biggr]
\,.\end{align}

\subsection{Fourier transform}
\label{app:Fourier}

The Fourier transform required when going from \eq{Iij_bare} to \eq{Iij_ren} can be conveniently evaluated using
\begin{align} \label{eq:FT_qT}
 \int\!\!\frac{\df^{2-2\eps}\qt}{\frac12 \Omega_{1-2\eps} q_T^{-2\eps}}
      \, e^{\img \bt \cdot \qt} \biggl(\frac{\mu^2}{q_T^2}\biggr)^{\ell\eps}  \frac{\ln^n\kappa}{q_T^2}
 = e^{\ell \eps L_b} \Gamma(1-\eps)
   \frac{\df^n}{\df^n\eta}\bigg|_{\eta=0} e^{\eta (L_\omega - L_s)}
    \frac{\Gamma(\eta-\ell\eps) e^{2(\eta-\ell\eps)\gamma_E}}{\Gamma[1-(\eta-\ell\eps) - \eps]}
\,,\end{align}
where $\kappa = \tau q_T^2 / \omega$, $\ell$ is the loop-order, and we express all resulting logarithms in terms of
\begin{align}
 L_b = \ln\frac{b_T^2 \mu^2}{4 e^{-2\gamma_E}}
\,,\qquad
 L_\omega = \ln\frac{1/\tau}{\omega}
\,,\qquad
 L_s = \ln\frac{b_T^2 / \tau^2}{4 e^{-2\gamma_E}}
\,.\end{align}
In \eq{FT_qT} we divide by the angular factor $\frac12 \Omega_{1-2\eps} = \pi^{1-\eps} / \Gamma(1-\eps)$ and $q_T^{-2\eps}$ to account for the fact that $q_T$ is defined as the magnitude of the $(2-2\eps)$-dimensional vector, and that the associated $2-2\eps$-dimensional solid angle has already been integrated over in $\cI_{ij}$.

\subsection{Renormalization group equations}
\label{app:RGEs}

The beam function depends on the renormalization scale $\mu$ and the rapidity scale $\nu$,
and thus obeys two coupled RGEs~\cite{Chiu:2012ir,Li:2016axz}
\begin{align} \label{eq:RGEs}
 \mu \frac{\df}{\df\mu}{\tilde B_i(x,b_T,\mu,\nu/\omega)} &
 = \tilde\gamma_B^i(\mu,\nu/\omega)\, \tilde B_i(x,b_T,\mu,\nu/\omega)
\,,\nn\\
 \nu \frac{\df}{\df\nu}{\tilde B_i(x,b_T,\mu,\nu/\omega)} &
 = -\frac{1}{2}\tilde\gamma_\nu^i(b_T,\mu)\, \tilde B_i(x,b_T,\mu,\nu/\omega)
\,,\end{align}
where $\tilde\gamma_\nu^i$ is the so-called rapidity anomalous dimension~\cite{Chiu:2012ir}.
Its prefactor or $-1/2$ arises because $\tilde\gamma_\nu^i$ is defined as the $\nu$-anomalous definition of the soft function.

The beam anomalous dimension has the all-order form
\begin{equation}
 \tilde\gamma_B^i(\mu,\nu/\omega) = 2 \GammaC^i[\as(\mu)] \ln\frac{\nu}{\omega} + \tilde\gamma_B^i[\as(\mu)]
\,,\end{equation}
where $\GammaC^i(\as)$ and $\tilde\gamma_B^i(\as)$ are the cusp and the beam noncusp anomalous dimensions in the appropriate color representation $i=q$ or $i=g$, but are independent of the quark flavor.

The RGE for the matching kernel follows from \eqs{beam_OPE}{RGEs} together with the DGLAP equation
\begin{equation} \label{eq:DGLAP}
 \mu \frac{\df}{\df\mu} f_i(z,\mu)
 = 2 \sum_j \int_z^1\!\frac{\df z'}{z'}\, P_{ij}(z',\mu)\, f_j\Bigl(\frac{z}{z'}, \mu\Bigr)
\,.\end{equation}
It is given by
\begin{align} \label{eq:I_RGE}
 \mu \frac{\df}{\df\mu} \tilde\cI_{ij}\Bigl(z,b_T,\mu,\frac{\nu}{\omega}\Bigr)
&= \sum_k \int_z^1 \frac{\df z'}{z'}\, \tilde\cI_{ik}\Bigl(\frac{z}{z'},b_T,\mu,\frac{\nu}{\omega}\Bigr)
   \Bigl[ \tilde\gamma_B^i\Bigl(\mu,\frac{\nu}{\omega}\Bigr) \delta_{kj} \delta(1-z') - 2  P_{kj}(z',\mu) \Bigr]
\,.\end{align}

The rapidity anomalous dimension itself is governed by an RGE in $\mu$,
\begin{align}
 \mu \frac{\df}{\df\mu} \tilde\gamma^i_\nu(b_T,\mu) = -4 \GammaC^i[\as(\mu)]
\,,\end{align}
which can be solved as
\begin{align}
 \tilde\gamma^i_\nu(b_T,\mu) &
 = -4 \int_{b_0/b_T}^\mu \frac{\df\mu'}{\mu'} \, \GammaC^i[\as(\mu')]
 + \tilde\gamma_\nu^i[\as(b_0/b_T)]
\,.\end{align}
Here, $b_0 = 2 e^{-\gamma_E}$, and $b_0/b_T$ is a conventional boundary scale.
The coefficients of the boundary term $\tilde\gamma_\nu^i[\as(b_0/b_T)]$
are defined as the constants of the rapidity anomalous dimension, which we write as
\begin{align}
 \tilde\gamma_\nu^i(\as)
 = \sum_{n=0}^\infty \gamma_{\nu\,n}^i \Bigl(\frac{\as}{4\pi}\Bigr)^{n+1}
\,.\end{align}
This anomalous dimension appears in the eikonal limit in \eq{eikonal_limit},
and is related to the notation of \refcite{Li:2016ctv} by $\gamma_2^r = 2 \gamma_{\nu\,2}^i$,
where the color representation $i$ is implicit in $\gamma_2^r$.

\subsection{Structure of the beam function counterterm}
\label{app:counter_term}

The beam function counterterm can be predicted from \eq{RGEs} using
\begin{align} \label{eq:mu_RGE_qT}
 \frac{\df}{\df\ln\mu} \ln{\tilde Z_B^i(\eps,\mu,\nu/\omega)}
 = - \tilde\gamma_B^i(\mu,\nu/\omega)
 = -2 \GammaC^i[\as(\mu)] \ln\frac{\nu}{\omega} - \tilde \gamma_B^i[\as(\mu)]
\,.\end{align}
The all-order form of the counterterm is given by (see also \refcite{Becher:2009cu})
\begin{align} \label{eq:ZB_qT}
 \ln \tilde Z_B^i(\eps,\mu,\nu/\omega) = -\int\limits_0^{\as(\mu)} \!\! \frac{\df\alpha}{\beta(\alpha,\eps)} \biggl[
  2 \GammaC^i(\alpha) \ln\frac{\nu}{\omega} + \tilde \gamma_B^i(\alpha) \biggr]
\,,\end{align}
where $\beta(\alpha_s,\eps) = -2 \eps \alpha_s + \beta(\alpha_s)$ is the QCD beta function in $d=4-2\eps$ dimensions. Expanding \eq{ZB_qT} systematically in $\alpha$, we obtain the result through three loops as
\begin{align} \label{eq:ZB_qT_2}
 \ln \tilde Z_B^i(\eps,\mu,\nu/\omega) &=
 \frac{\as}{4\pi} \frac{1}{2\eps} \bigl( 2\Gamma_0^i L_\omega + \tilde\gamma_{B\,0}^i \bigr)
\nn\\&
 + \Bigl(\frac{\as}{4\pi}\Bigr)^2 \Bigl[
   - \frac{\beta_0}{4 \eps^2} \bigl( 2\Gamma_0^i L_\omega + \tilde\gamma_{B\,0}^i \bigr)
   +\frac{1}{4 \eps} \bigl( 2\Gamma_1^i L_\omega + \tilde\gamma_{B\,1}^i \bigr)
 \Bigr]
\nn\\&
 + \Bigl(\frac{\as}{4\pi}\Bigr)^3 \biggl\{
   \frac{\beta_0^2}{6\eps^3} \bigl( 2\Gamma_0^i L_\omega + \tilde\gamma_{B\,0}^i \bigr)
   - \frac{1}{6\eps^2} \bigl[ \beta_1 \bigl( 2\Gamma_0^i L_\omega + \tilde\gamma_{B\,0}^i\bigr) + \beta_0 \bigl( 2\Gamma_1^i L_\omega + \tilde\gamma_{B\,1}^i \bigr) \bigr]
   \nn\\&\hspace{1.8cm}
   +\frac{1}{6\eps} \bigl( 2 \Gamma_2^i L_\omega + \tilde\gamma_{B\,2}^i \bigr)
 \biggr\}
 + \cO(\as^4)
\,.\end{align}
Here, $L_\omega = \ln(\nu/\omega)$, and the $\gamma_n$ are the coefficients of the corresponding anomalous dimensions at $\cO[(\as/4\pi)^n]$. Explicit expressions for all anomalous dimensions in the convention of \eq{ZB_qT_2} are collected in~\refcite{Billis:2019vxg}.
The required three-loop results for $\GammaC$ and $\beta$ were calculated in \refscite{Korchemsky:1987wg, Moch:2004pa, Vogt:2004mw} and \cite{Tarasov:1980au, Larin:1993tp}, respectively.
The coefficients of $\tilde\gamma_B$ follow from consistency with the anomalous dimensions of the hard and soft functions in \eq{TMD_factorization}, which can be obtained from the quark and gluon anomalous dimensions of the corresponding form factors calculated in~\refscite{Kramer:1986sg, Matsuura:1987wt, Matsuura:1988sm, Harlander:2000mg, Gehrmann:2005pd, Moch:2005id, Moch:2005tm}.
Our calculation explicitly confirms the beam anomalous dimension obtained from these relations.

\subsection[\texorpdfstring{$\as$}{alphaS} renormalization and IR counterterms]
           {\boldmath $\as$ renormalization and IR counterterms}
\label{app:UV_IR_counter_terms}

The bare strong coupling constant $\as^b$ is renormalised as
\bea \label{eq:asb}
\as^b&=&\as\left(\frac{\mu^2}{4\pi} e^{\gamma_E}\right)^{\eps} \left[1+\frac{\as}{4\pi}\left(-\frac{\beta_0}{\epsilon }\right)+\left(\frac{\as}{4\pi}\right)^2\left(\frac{\beta_0^2}{\epsilon ^2}-\frac{\beta_1}{2 \epsilon }\right)\nn\right.\\
&&\hspace{2.5cm}+\left.\left(\frac{\as}{4\pi}\right)^3\left(-\frac{\beta_0^3}{\epsilon ^3}+\frac{7 \beta_1 \beta_0}{6 \epsilon ^2}-\frac{\beta_2}{3 \epsilon }\right)+\mathcal{O}(\as^4)\right]\,.
\eea
The mass factorisation counter term can be expressed in terms of the splitting functions~\cite{Moch:2004pa,Vogt:2004mw} as
\bea \label{eq:Gamma}
\Gamma_{ij}(z) =&& \delta_{ij}\delta(1-z) \nn\\
&&+\left(\frac{\as}{4\pi}\right)\frac{P^{(0)}_{ij}}{\epsilon} \nn\\
&&+\left(\frac{\as}{4\pi}\right)^2\left[\frac{1}{2\epsilon^2}\left(P^{(0)}_{ik}\otimes P^{(0)}_{kj}-\beta_0  P^{(0)}_{ij}\right)+\frac{1}{2\epsilon}P^{(1)}_{kj}\right]\nn\\
&&+\left(\frac{\as}{4\pi}\right)^3\left[\frac{1}{6\epsilon^3}\left(P^{(0)}_{ik}\otimes P^{(0)}_{kl}\otimes P^{(0)}_{lj}-3\beta_0P^{(0)}_{ik}\otimes P^{(0)}_{kj}+2\beta_0^2P^{(0)}_{ij}\right)\right.\nn\\
&&\hspace{1.6cm}\left.+\frac{1}{6\epsilon^2}\left(P^{(1)}_{ik}\otimes P^{(0)}_{kj}+2P^{(0)}_{ik}\otimes P^{(1)}_{kj}-2\beta_0 P^{(1)}_{ij}-2 \beta_1 P^{(0)}_{ij}\right)+\frac{1}{3\epsilon}P^{(2)}_{ij}\right].\nn\\
&&+~\mathcal{O}(\as^4)
\,.\eea
Here, we suppress the argument $z$ of the splitting functions on the right hand side and keep the summation over repeated flavor indices implicit. The convolution in \eq{Gamma} is defined as
\beq
f\otimes g=\int_z^1\frac{\df z^\prime}{z^\prime} f(z)g\left(\frac{z}{z^\prime}\right).
\eeq

\section{High-energy limit of the beam function kernels}
\label{app:high_energy}

The high-energy limit $z\to0$ of the kernels $\tilde I_{gg}^{(3)}(z)$ and $\tilde I_{gq}^{(3)}(z)$ contributing to the gluon beam function is given by
{
\allowdisplaybreaks
\bea
\lim\limits_{z\rightarrow 0}  z \, \tilde I_{gg}^{(3)}(z)&=&
C_A^3 \zeta_3\ln^2(z)+\left[C_A^3 \left(-\frac{469\zeta_2}{108} -\frac{11 \zeta_3}{12}-\frac{49 \zeta_4}{24}+\frac{1181}{81}\right)
\right.\nn\\&+&\left.
C_A^2 n_f \left(-\frac{4\zeta_2}{27} +\frac{5 \zeta_3}{6}+\frac{49}{324}\right) +C_A C_F n_f \left(\frac{8}{27}  \zeta_2- \zeta_3-\frac{311 }{486}\right)\right] \ln(z) 
\nn\\&+&
n_f C_F^2  \left(\frac{5\zeta_2}{36} -\frac{7\zeta_3}{9} +\frac{8\zeta_4}{9} +\frac{3 }{32}\right)-C_A n_f^2 \left(\frac{\zeta_3}{18}+\frac{1255}{5832}\right)
\nn\\&+&
C_A^3 \left(2 \zeta_3 \zeta_2-\frac{3529 \zeta_2}{162}-\frac{28 \zeta_3}{3}-\frac{77 \zeta_5}{4}-\frac{363 \zeta_4}{16}+\frac{1572769}{15552}\right)
\nn\\
&+&C_A^2 n_f \left(-\frac{509 \zeta_2}{324} +\frac{65 \zeta_4}{72}-\frac{4 \zeta_3}{9}+\frac{66881}{11664}\right) +n_f^2 C_F \left(\frac{559}{2916}-\frac{2}{9} \zeta_3\right)  
\nn\\&+&
C_A C_F n_f \left(\frac{317 \zeta_2}{108} +\frac{2\zeta_3}{9} -\frac{43\zeta_4}{36} -\frac{418097 }{46656}\right)
\,,\\
\lim\limits_{z\rightarrow 0} z \, I_{gq}^{(3)}(z)&=&
  C_A^2 C_F \zeta_3 \ln^2(z)+ \left[ C_F C_A^2\left(\frac{469}{108} \zeta_2-\frac{5}{4} \zeta_3-\frac{49}{24} \zeta_4+\frac{1181}{81}\right) 
\right. \nn\\ &+& \left.
C_F^2 n_f \left(\frac{8\zeta_2 }{27}  -\frac{\zeta_3}{3}-\frac{595}{972}\right) +C_F C_A n_f \left(-\frac{4\zeta_2}{27} +\frac{\zeta_3}{2} +\frac{89 }{648}\right)\right] \ln(z)
\nn\\
&+& C_F C_A^2 \left(-\frac{12913}{648} \zeta_2-\frac{47}{6} \zeta_3+\frac{5}{3} \zeta_2 \zeta_3-\frac{3109}{144} \zeta_4-\frac{263}{12} \zeta_5+\frac{333613}{3456}\right)
\nn\\
&+& C_F^2 C_A \left(\frac{11 }{8} \zeta_2 -\frac{21}{2} \zeta_3-\frac{8}{3} \zeta_2 \zeta_3+\frac{23}{4} \zeta_4+\frac{38}{3} \zeta_5-\frac{1105}{384}\right)
\nn\\
&-&C_F n_f^2 \left(\frac{\zeta_3}{6}+\frac{29}{243}\right) + C_F^2n_f \left(\frac{611 \zeta_2}{648}-\frac{\zeta_4}{4}-\frac{53 \zeta_3}{36}-\frac{69667}{46656}\right)
\nn\\
&+& C_F C_A n_f \left(-\frac{19 }{27} \zeta_2+\frac{13}{18} \zeta_3+\frac{\zeta_4}{8}+\frac{17137}{7776}\right)
\nn\\
&+& C_F^3 \left(3 \zeta_3 \zeta_2-\frac{13 \zeta_2}{8}+\frac{37 \zeta_3}{4}-10 \zeta_5-\frac{49 \zeta_4}{8}+\frac{467}{192}\right)
\,.\eea
}%
Here, the color factors $C_A$ and $C_F$ are only used for compactness of the result and should be replaced with their expressions in terms of $n_c$.
The corresponding limit for the quark kernels were already presented in \refcite{Luo:2019szz}, for which we find perfect agreement.
Note that the expressions for the high energy limit ${z\rightarrow 0}$ up to $\cO(z^{50})$, as well as that for the threshold limit $z\to 1$ up to $\cO((1-z)^{50})$, can be found for all channels in electronic form in the ancillary files of this work.

\addcontentsline{toc}{section}{References}
\bibliographystyle{jhep}
\bibliography{../refs}

\end{document}